\documentclass[apjl]{emulateapj}

\newcommand{\kms}{km s$^{-1}$}
\newcommand{\magsec}{mag arcsec$^{-2}$}

\newcommand{\dw}{$\Delta w$}
\newcommand{\tmax}{$t_{max}$}
\newcommand{\lcdm}{$\Lambda$CDM}
\newcommand{\eg}{e.g.,\ }
\newcommand{\etal}{et~al.\ }
\newcommand{\Msun}{M$_{\sun}$}

\received{2009 January 23}
\accepted{2009 May 7}

\begin{document}

\title{Tidal Streams of Intracluster Light}
 
\author{Craig S. Rudick, J. Christopher Mihos, Lucille
H. Frey\altaffilmark{1}}
\email{csr10@case.edu, mihos@case.edu, lfrey@lanl.gov}
\affil{Department of Astronomy, Case Western Reserve University, 10900
Euclid Ave, Cleveland, OH 44106} \altaffiltext{1}{Now at Los Alamos
  National Laboratory, Los Alamos, NM, USA.}

\and
\author{Cameron K. McBride}
\email{cameron.mcbride@vanderbilt.edu}
\affil{Department of Physics \& Astronomy, Vanderbilt University, 6301
  Stevenson Center, Nashville, TN 37235}

\begin{abstract}
Using $N$-body simulations, we have modeled the production and
evolution of substructures in the intracluster light (ICL) of a
simulated galaxy cluster.  We use a density-based definition of ICL,
where ICL consists of luminous particles which are at low densities,
to identify ICL particles and track their evolution.  We have
implemented a friends-of-friends-type clustering algorithm which finds
groups of particles correlated in both position and velocity space to
identify substructures in the ICL, hereafter referred to as
``streams''.  We find that $\approx 40\%$ of the cluster's ICL is
generated in the form of these massive ($M\ge7.0\times10^{8}$ \Msun),
dynamically cold streams.  The fraction of the ICL generated in
streams is greater early in the cluster's evolution, when galaxies are
interacting in the group environment, than later in its evolution when
the massive cluster potential has been assembled.  The production of
streams requires the strong tidal fields associated with close
interactions between pairs of galaxies, and is usually associated with
merging pairs of galaxies, or fast, close encounters with the
cluster's central galaxy.  Once streams are formed, they begin to
decay as they are disrupted by the tidal field of the cluster.  We
find that streams have decay timescales which are $\approx 1.5$ times
their dynamical time in the cluster.
\end{abstract}


\section{Introduction}
Diffuse intracluster starlight (ICL) consists of stars in galaxy
clusters which are observed to be outside of and distinct from the
cluster galaxies.  In the prevailing \lcdm\  hierarchical mass assembly
paradigm, the ICL is generated from the numerous gravitational
interactions experienced by the cluster galaxies as they interact with
other galaxies and the cluster potential during the cluster's
formation and evolution.  As a product of the dynamical interactions
within the cluster, the ICL has the potential to reveal a great deal
of information about the cluster's accretion history and evolutionary
state.  The quantity, morphology, and kinematics of the ICL each hold
useful information related to the evolution of the cluster, as well as
its constituent galaxies.

Observationally, ICL has been detected in numerous galaxy clusters
through broadband imaging (\eg Zwicky 1951; Uson \etal 1991;
V{\'i}lchez-G{\'o}mez et al. 1994; Gonzalez \etal 2005; Mihos \etal
2005; Zibetti et al. 2005; Krick \& Bernstein 2007) as well as the
detection of discrete tracers, such as planetary nebulae (\eg
Feldmeier \etal 1998; Arnaboldi et al. 2004), red giants (\eg Ferguson
\etal 1998; Durrell \etal 2002; Williams \etal 2007a), and other types
of stellar tracers (\eg Neill \etal 2005; Gal-Yam et al. 2003;
Williams \etal 2007b).  Most studies estimate that the ICL comprises
$\approx 10\%$ -- $30\%$ of the clusters' luminosity.  While it is now
thought that ICL is a ubiquitous feature of evolved galaxy clusters,
it is likely that there is no universal ICL fraction, but that
different clusters will have different ICL contents, depending on its
specific evolution and history (Murante \etal 2004; Conroy \etal 2007;
Rudick \etal 2006, hereafter R06).

A number of observational studies have also shown that there is often
distinct substructure found in the morphology of the ICL.  These
features can take numerous forms, including long straight streamers
(Mihos \etal 2005), curved arcs (Trentham \& Mobasher 1998;
Calc{\'a}neo-Rold{\'a}n \etal 2000), tidal tails (Krick \etal 2006),
large plumes (Gregg \& West 1998; Feldmeier \etal 2004), and bridges
spanning galaxies (Feldmeier \etal 2002).  However, due to the
difficulty in quantifying the morphological characteristics of such
substructures, descriptions of the their morphologies remain largely
qualitative and imprecise.

While the existence of ICL is now well established, the processes
which generate it are not well understood.  There are a number of
processes, however, which have been shown to be capable of stripping
material from galaxies, including stripping during the initial
collapse of the cluster (\eg Merrit 1984); stripping of galaxies by an
established cluster potential (Byrd \& Valtonen 1990; Gnedin 2003);
stripping within galaxy groups accreting onto the cluster (Mihos 2004,
R06); and stripping from high speed encounters between cluster
galaxies (Moore et al. 1996).  In the complex environment of a
collapsing and accreting galaxy cluster, all these processes likely
contribute to the overall production of ICL.

Several recent studies have suggested that galaxy mergers are a
primary mechanism responsible for generating ICL (Monaco \etal 2006;
Stanghellini \etal 2006; Conroy \etal 2007; Murante \etal 2007).  Such
merger processes leading to the creation of a diffuse stellar
component are acting not only in massive galaxy clusters, but are also
present in the hierarchical build-up of less massive structures, such
as groups (Sommer-Larsen 2006) and even individual galaxy halos
(Purcell \etal 2007).  Several observational studies have seen
evidence for diffuse, intergalactic light in galaxy groups (Nishiura
\etal 2000; White \etal 2003; Da Rocha \& Mendes de Oliveira 2005;
Aguerri \etal 2006; Da Rocha \etal 2008), often involving significant
substructure.  On galactic scales, the substructure seen in the Milky
Way's stellar halo (\eg Harding \etal 2001; Johnston \etal 2008;
Belokurov \etal 2006; Juri{\'c} \etal 2008) indicates that much of the
Galaxy's stellar halo is the product of tidal disruption of satellite
galaxies, analogous to the production of ICL in galaxy clusters.

In order to study to generation of ICL in clusters more directly, a
number of recent studies have focused on simulating cluster evolution
within a cosmological volume, where the ICL is traced by dark matter
simulations tracer particles (\eg Napolitano \etal 2003),
collisionless galaxy models (R06), or full hydrodynamical models (\eg
Murante \etal 2004; Willman \etal 2004; Sommer-Larsen \etal 2005).
These simulations have generally found that at $z=0$, at least 10\% of
the clusters' stars were found in the diffuse ICL, in line with
current observations.  However, a direct comparison between the
results of simulations and observations is complicated by the varying
metrics used to define ICL in observations and simulations (see Section
\ref{sec:density}).

By combining both positional and kinematic data, we have the potential
to dramatically increase our understanding of the ICL.  Napolitano
\etal (2003), Sommer-Larsen \etal (2005), and Willman \etal (2004)
each study the kinematic distribution of the unbound stars and find
significant kinematic substructure.  Observationally, planetary
nebulae studies have successfully measured line-of-sight velocities of
intracluster stars in several galaxy clusters, including the Virgo
(Arnaboldi \etal 2004) and Coma (Gerhard \etal 2005) clusters.
Further studies, which include larger sample sizes, will allow the
study of ICL substructure in phase space, many which may not
detectable in position space alone.

In R06 we used a set of $N$-body simulations of galaxy clusters to
measure the evolution of the ICL from an observational perspective.
In this paper, we use one of these simulations to objectively identify
and measure the evolution of substructure within the ICL.  Section 2
gives an overview of the $N$-body simulation technique.  In Section 3,
we define a density-based definition of ICL, and describe the
algorithm we have developed to identify ICL substructures; these
substructures are referred to as ``streams'' throughout this paper.
Section 4 measures the prevalence of streams within the ICL, while
Sections 5 and 6 follow the evolution of individual galaxies
and streams, respectively, in order to better understand the
production and evolution of the streams.  Finally, Section 7 contains
a summary of our findings and a discussion of their implications.

\section{The Simulation}

The simulation used to study the intracluster light in this paper was
first described in R06; we reiterate only a brief outline of the
simulation process here.

Our simulation technique begins by running a $N=256^3$ $50 \times 50 \
\times 50$ Mpc $\Lambda=0.7$, $\Omega_M=0.3$, $H_0=70$ \kms\
Mpc$^{-1}$ cosmological dark matter simulation from $z=50$ to $z=0$.
From this simulation, we identify massive dark matter halos, with
masses $`10^{14}$ \Msun, to re-simulate at higher resolution.  For
each cluster, we identify the $z=2$ halos which will later constitute
the $z=0$ cluster halo.  Into these $z=2$ halos we insert higher
resolution collisionless galaxy models, substituting the most bound
70\% of the original halo mass and leaving the remaining 30\% to form
an extended dark matter halo around the galaxy.  For high-mass halos,
we employ a ``halo occupancy distribution'' (HOD) technique (Berlind
\& Weinberg 2002), which allows us to populate single halos with
multiple galaxy models; we also have a minimum halo mass, below which
we do not insert a galaxy model.  We use two types of galaxy models,
initialized using the prescriptions of Hernquist (1993): a disk model
in which the stars follow a composite exponential disk plus Hernquist
(1990) bulge (with bulge-to-disk ratio of 1:5), and an elliptical
galaxy model where the stars have a pure Hernquist (1990)
distribution.  The luminous particles which make up these galaxy
models, and on which our analyses focus, have a mass of $1.4 \times
10^{6}$ \Msun, with a gravitational softening length of 280 pc.  Both
models are embedded in isothermal dark halos consisting of particles
of intermediate mass resolution, lower than luminous matter particles
but higher than the original dark matter particles which make up the
extended halo. The composite dark matter profile resulting from the
particles in both the isothermal halo and the original extended dark
matter distribution yields a flat rotation curve in the luminous disk,
and exhibits a $r^{-3}$ Navarro-Frenk-White (NFW)-like decline in
density at large radius.  Our galaxy models are scaled to match the
mass of the substituted halo, and our galaxies have a total
dark-to-luminous mass ratio of 10:1.  Our final galaxy mass function
behaves like a Schecter function with a high mass cutoff and at low
masses shows a power law slope, $\alpha \approx -1$.  With the $z=2$
initialization complete, the cluster is evolved to $z=0$ using the
$N$-body code GADGET (Springel et al. 2001).

This technique allows us to study the detailed structure and evolution
of the ICL in massive galaxy clusters in a cosmological environment,
while keeping the simulations computationally feasible.  We recognize,
however, a number of caveats which need to be addressed when
interpreting these results.  Our collisionless galaxy models focus
entirely on gravitational dynamics, while neglecting the various
effects of hydrodynamics.  We are thus isolating gravitational
stripping as the mechanism driving ICL production, irrespective of the
effects of star formation, ram pressure stripping, or other
hydrodynamic processes.  Also, in our galaxy substitution scheme, we
have a minimum galaxy mass corresponding to $6 \times 10^{10}$ \Msun.
For a Shechter luminosity function with faint end slope $\alpha=-1$,
galaxies below this minimum mass contain about 10\% of the cluster's
total luminosity; we are unable to measure these galaxies'
contribution to the ICL.  Finally, we initialize our galaxy models
into the dark matter halos which exist at $z=2$.  Although we use a
mix of both elliptical and disk galaxy models, each initialized galaxy
is in an identical evolutionary state (although scaled in mass) as
each other galaxy of the same type, ignoring the heterogeneity caused
by the complexities of early galactic evolution.  While most ICL
formation is expected to take place at $z \leq 1$ (\eg R06; Murante
\etal 2007), any ICL formed at $z \geq 2$ (\eg Saro \etal 2009) will be
missed by our simulation technique.  Our initialization at $z=2$ also
means that we only place galaxies into halos which meet our minimum
halo mass criterion at that time, and not halos which only
subsequently grow to reach that mass.

In this paper, we focus on a detailed analysis of one of one such
cluster (cluster C2 from R06).  At $z=0$, the cluster has $R_{200} =
917$ kpc (the radius within which the density of the cluster is 200
times the critical density) and $M_{200} = 8.4 \times 10^{13}$ \Msun\
(the mass enclosed within $R_{200}$).  The cluster is initialized with
144 galaxies, while only 95 remain at $z=0$ due to galaxy mergers (see
Section \ref{sec:streams} for a more detailed discussion of mergers).  50
galaxies are inside of $R_{200}$ at $z=0$, containing 75\% of the
stellar mass in the simulation.

\begin{figure*}
\plotone{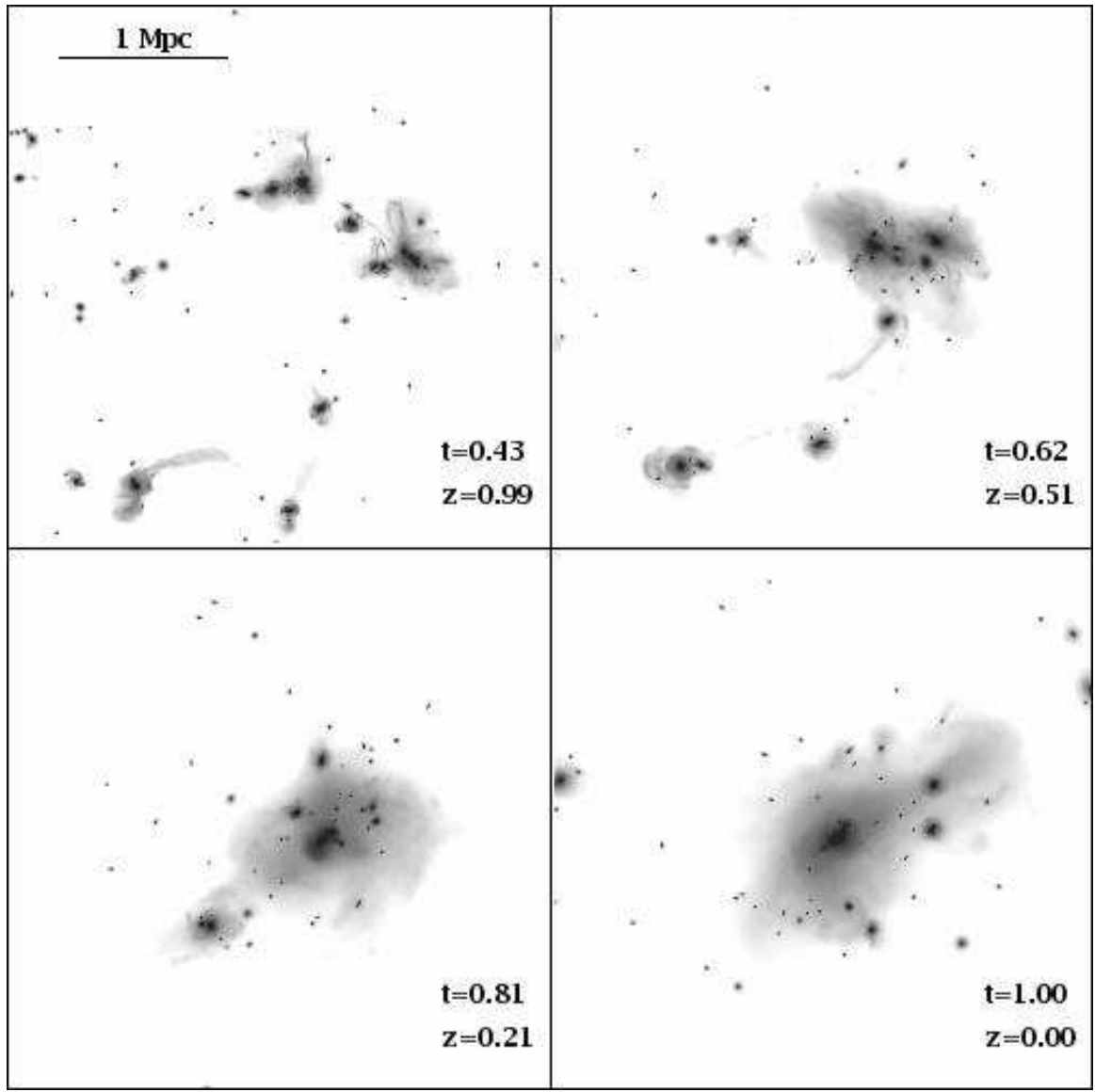}
\caption{The cluster at four stages during its evolution.  The
  gray-scale color is smoothed mass surface density of the
  cluster's luminous matter.  There is no calibration of the gray-scale
  color, but is meant to show only relative surface density, and was
  chosen to highlight important features of the cluster.  The physical
  scale of all four images is identical, and shown by the bar in the
  upper left of the figure.  Each image is labeled by the time (in
  units of fraction of the age of the universe), $t$, and redshift,
  $z$, in the lower right.
\label{fig:cluster_image}}
\end{figure*}

Figure \ref{fig:cluster_image} shows the evolution of this cluster,
showing the smoothed surface mass density distribution of the
cluster's luminous matter at four timepoints, roughly equally spaced
in evolutionary time.  The images were made in the same manner as
those shown in R06, however here we have not converted the surface
mass density into a surface brightness.  The hierarchical assembly of
the cluster can be clearly seen, and is of extraordinary importance in
understanding all aspects of this cluster's evolution, including its
ICL and ICL streams.  Early in the cluster's history, at
$t=0.43$\footnote{Unless otherwise specified, throughout this paper
all times are given as a fraction of the age of the universe at
$z=0$.}, there is not yet a central cluster, just a number of galaxy
groups.  The galaxies within these groups are merging and interacting
with one another, but the groups themselves are still relatively
independent of one another.  What low surface density material
(equivalent to the low surface brightness ICL from R06) there is, is
concentrated around the small groups, and displays a number of long,
linear features.  As the cluster evolves, the galaxy groups begin to
merge together to form the cluster.  By $t=0.62$, a number of groups
have begun to merge and interact with one another, and by $t=0.81$, a
cluster core is clearly visible.  The low surface density material has
grown in extent around the cluster core, many of the linear structures
seen at earlier evolutionary times have been destroyed by the cluster
potential.  These trends are continued until $t=1$, where we see that
the cluster has a central core with a large cD galaxy at its center,
with a large extended envelope of low surface mass density material
surrounding the core.  On the far edges of the $t=1$ image, several
examples of galaxies which are well outside the cluster and have not
yet had any significant interactions with the cluster potential can be
seen.  Understanding this hierarchical evolution will be extremely
important when interpreting the evolution of the ICL and streams
discussed in this paper.

\section{Defining ICL and Streams}

\subsection{ICL Defined by Density} \label{sec:density}

There is no standard definition of ICL, and many authors have used
different metrics to define and measure ICL in the literature.  Common
observational definitions of ICL include luminosity in excess of a
fitted central galaxy profile (\eg Gonzalez \etal 2005) or luminosity
fainter than a given surface brightness limit (\eg Feldmeier \etal
2004).  In R06 we created simulated surface brightness maps of our
clusters and measured the ICL content using the latter definition.
However, any observationally tractable definition of ICL suffers from
projection effects, and is therefore not well linked to the cluster
physics.  Most studies of ICL using numerical simulations have used a
binding energy definition of ICL; i.e. ICL consists of stars which are
not bound to any individual galaxy in the cluster potential (\eg
Napolitano \etal 2003; Murante \etal 2004; Sommer-Larsen \etal 2005).
While this is an appealing definition from a theoretical perspective
--- since all stars are initially bound to their parent galaxy,
unbound stars must have had physical forces act on them to change
their binding energy --- in practice, calculating the binding energy
of particles requires making numerous assumptions about the mass
distribution of the cluster and its galaxies.  This can become
especially difficult and convoluted during the complex merging
processes which dominate the cluster's evolution at early times.

In this paper we have chosen to adopt a density-based definition of
ICL which can be easily defined, repeatably measured, and intuitively
understood throughout the cluster's evolution.  The simplest form such
a density-based definition of ICL would be to simply define a limiting
density threshold, $\rho_{thresh}$, below which particles are
considered ICL.  We use a slightly modified version of this definition
whereby we require ICL particles to remain below the density threshold
for a minimum consecutive time period, $t_{min}$.  This simple
addendum helps to eliminate loosely bound particles on radial orbits
around their host galaxy from being defined as ICL.  Additionally,
once we have identified a particle as ICL, it remains an ICL particle
for the rest of the simulation, irrespective of its future density.
This scheme for identifying ICL is premised on the simple model of ICL
formation whereby particles are pulled from high to low density as
they are stripped from their parent galaxies, and are thereafter
unbound particles orbiting the cluster potential.  We find that only a
very small fraction of the particles we identify as ICL later become
rebound to galaxies and are found in the high-density galactic cores,
and do not affect the streams described in this paper.

\begin{figure*}
\plotone{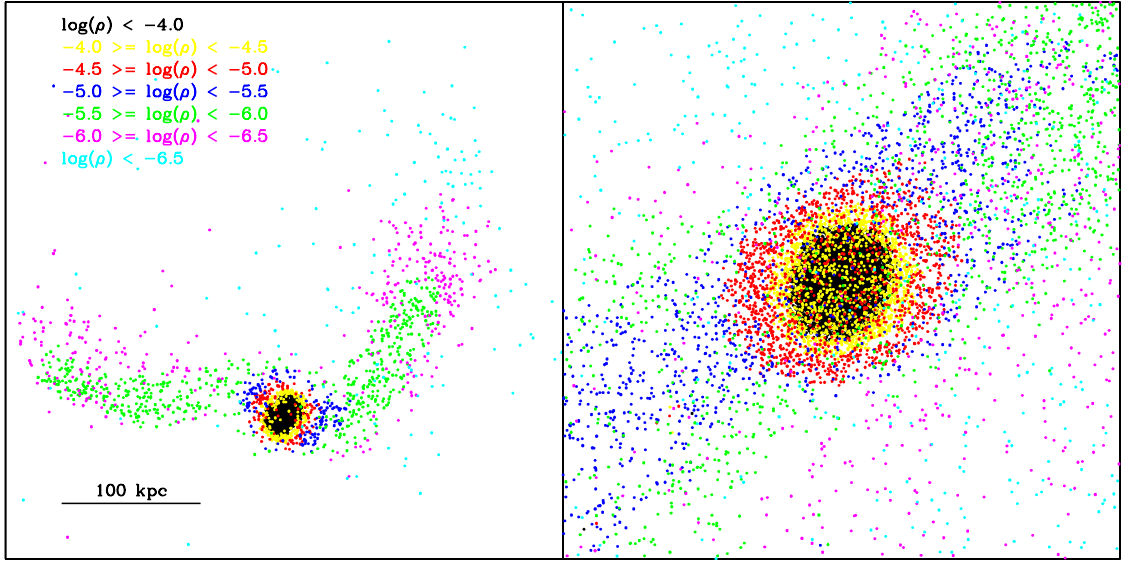}
\caption{
\label{fig:stream_density}
Two-dimensional projections of two galaxies at $z=0$, showing the
density of their luminous particles in various colors, as labeled by
the key in the upper left.  All densities are in units of \Msun\
pc$^{-3}$.  The galaxy on the left is a low-mass disk galaxy, and the
galaxy on the right is a high-mass elliptical merger product.  Both
galaxies are plotted on the same physical scale, as shown in the lower
left.  For clarity, only 10\% of the particles are shown for the
galaxy on the right.
}
\end{figure*}

Our ICL definition requires us to define a density threshold,
$\rho_{thresh}$, below which particles are considered ICL.  This
choice is very similar to that faced in R06 where we defined ICL as
luminosity below a given surface brightness threshold - namely, that
the threshold choice is essentially arbitrary, but because we are
defining ICL using a well-defined property, the measurement is
extremely repeatable (see the end of Section \ref{sec:streams} for more on
the connections between density and surface brightness).  As in R06,
we have chosen a threshold value that attempts to select only what we
have identified as ICL by eye and intuition.  Figure
\ref{fig:stream_density} shows the density of particles from two
example galaxies at z=0.  Because we are primarily interested in tidal
streams, we have chosen a threshold density of $\rho_{thresh}=10^{-5}$
\Msun\ pc$^{-3}$, which we feel best separates the luminosity of the
galaxy from that of the tidal streams clearly present in these
galaxies.  While there is no single obvious value of $\rho_{thresh}$
which is ideal for all galaxies in all environments, we find that
higher values tend to classify too many particles in the outskirts of
galaxies as ICL, whereas lower values miss particles which are clearly
important contributors to the ICL.  The density of each particle is
defined as the mass density of luminous particles within the distance
to that particle's 100-th nearest neighbor.

Our ICL definition also contains minimum time, $t_{min}$, to ensure
that we do not label particles as ICL which are only in a short,
transitory low-density state as they orbit in the outskirts of a
galaxy.  We have set $t_{min} = 200$ Myr, which
is comparable to the dynamical timescale of the galaxies in our
simulations.  We feel that this criterion makes our definition more
robust, and adding the minimum time criterion reduces the number of
particles identified as ICL by only $\approx 10\%$ -- $15\%$ and does not
have a significant impact on our conclusions.  In order to determine
whether particles which become ICL near the end of the simulation
satisfy the $t_{min}$ criterion, we have evolved our simulation for a
short time past $z=0$ in order to measure the particle densities.

\begin{figure}
\epsscale{1.2}
\plotone{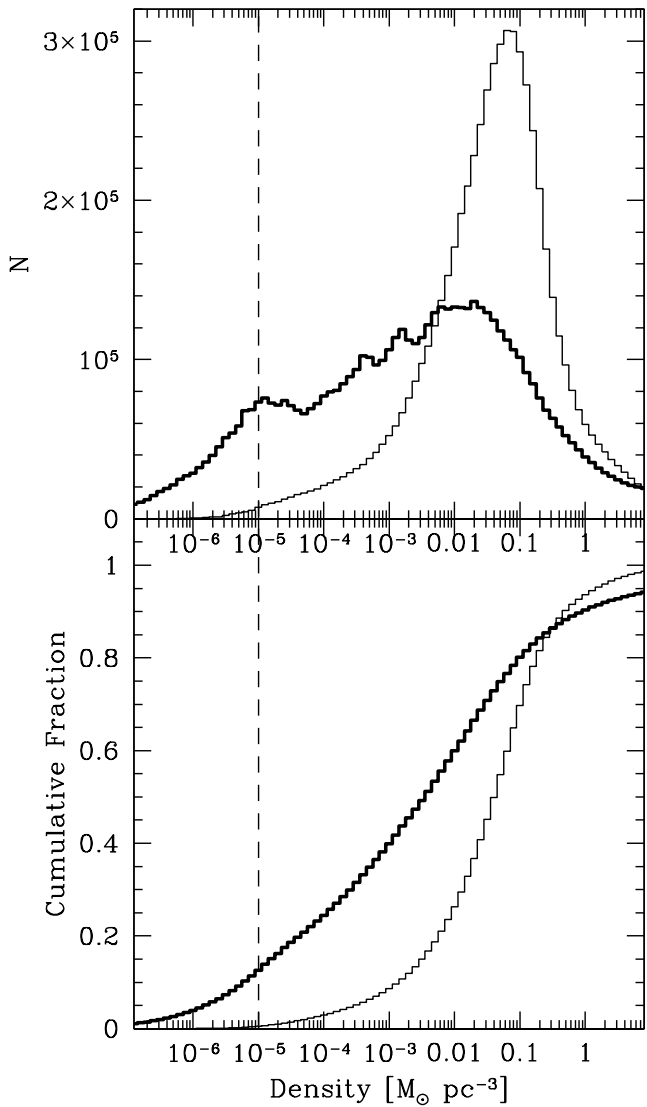}
\caption{
  Top: the histogram of the number of luminous particles as a function
  of density at $z=2$ (thin line) and $z=0$ (thick line).  Bottom: the
  cumulative histogram of the fraction of luminous particles at or
  above the given density at $z=2$ (thin line) and $z=0$ (thick line).
  The vertical dashed line indicates the value of $\rho_{thresh}$.
\label{fig:density_hist}}
\end{figure}

We find that a very small fraction of the luminous mass of our
galaxies, $\approx 0.2\%$, is classified as ICL when we initialize the
cluster at $z=2$, before any evolution has taken place.  This material
is, for the most part, comprised of loosely bound particles on the
outskirts of the most massive elliptical galaxies.  However, Figure
\ref{fig:density_hist}, which shows the distribution of particle
densities at $z=0$ and $z=2$ for cluster C2, clearly shows that this
is a minor correction and that this contribution is dwarfed by the ICL
that is generated during the evolution of the cluster. Figure 3 also
shows that our choice of $\rho_{thresh}=10^{-5}$ corresponds to what
appears to be a secondary peak in the $z=0$ density distribution.
While it is possible that this peak is the result of a discrete
population of stellar particles at low density corresponding to the
ICL, we have examined the particle distributions as a function of
density and found no evidence that this is the case.  The value of
$\rho_{thresh}$ was chosen based on an examination of tidal structures
on galactic scales as described above, and is independent of the
details of this secondary peak in the density distribution.

Thus, although any ICL definition includes trade-offs, we feel that
our density-based definition is advantageous in that it is a simple,
rigorous measurement, containing only well understood caveats.  One
downside to both a density and a binding energy definition of ICL is
that neither is observationally tractable.  The calculation of density
requires full three-dimensional position information for all ICL
stars, and binding energy requires full three-dimensional velocity
information, plus a detailed knowledge of the dark matter mass
distribution of galaxies.  However, a density-based definition of ICL
has the advantage of being analogous and closely related to the
observationally tractable surface density or surface brightness
definition of ICL.

\subsection{ICL Streams} \label{sec:streams}

We have identified ICL streams as groups of ICL particles which are
clustered in phase space.  The clustering algorithm used to detect the
streams is based on the DBSCAN algorithm (Ester \etal 1996) and is
essentially a friends-of-friends (FOF) spatial clustering algorithm
with an added free parameter which is used to reduce the inclusion of
``noise'' points from the clusters.  Whereas a traditional FOF
algorithm links together all points separated by less than a
pre-defined linking length, $\ell$, the DBSCAN algorithm first
separates the data into ``core'' and ``border'' points based on the
local density.  Core points are high density points with $k$ or more
data points whose distance is less than $\ell$; border points have
fewer than $k$ points with distance less than $\ell$.  The algorithm
then proceeds in much the same way as a FOF algorithm, linking points
separated by a distance less than $\ell$, except that clusters are
made up of only linked core points and the border points which are
linked to these core points.  Essentially, the DBSCAN algorithm is
more restrictive than a FOF algorithm by eliminating the connections
between border points.  Generally, this can have the effect of
reducing the size of clusters as compared with FOF, especially near the
cluster's borders where the local density may be lower, or of
splitting FOF clusters into multiple pieces by eliminating the border
point connections between clusters.  Note that by setting $k=1$, the
DBSCAN algorithm is identical to FOF.  In both the FOF and DBSCAN
algorithms, all points which are not linked to a cluster are
considered ``noise'' points and do not belong to any cluster.  The
number of clusters is not pre-determined, and clusters can be of
arbitrary shape, containing any number of data points greater than
$k$.

In both the FOF and DBSCAN algorithms, the distance between points
need not be the Euclidean distance, but can be any suitable metric, so
long as the distance between each pair of points is well-determined.
Because we are looking for tidal streams resulting from galaxy
interactions as the galaxy cluster evolves, we expect that the
luminous particles that we are seeking to cluster will be correlated
in both position and velocity space.  We have thus defined a phase
space distance metric, \dw, that combines both position and velocity
information to determine the phase space distance between a pair of
points:
\begin{equation} \Delta w = \sqrt{ \left( \frac{|\Delta
      \vec{x}|}{x_{n}}\right)^2 + g \left( \frac{|\Delta
      \vec{v}|}{v_{n}}\right)^2 }
\end{equation} where $|\Delta \vec{x}|$ is the distance in position
space and $|\Delta \vec{v}|$ is the distance in velocity space between
the pair; $x_{n}$ and $v_{n}$ are normalization factors for the
position and velocity, respectively; and $g$ is used as a weighting
factor to change the relative contributions of the position and
velocity terms.  We have set the values of the normalization factors
to be $x_n = 10$ kpc and $v_n = 30$ km s$^{-1}$, since these are scales
characteristic to kinds of tidal streams for which we are searching
(\eg Hibbard \& Mihos 1995).  However, note that these normalizations
merely provide a convenient starting point for choosing the values of
$g$ and $\ell$, which, along with $k$, fully determine the clustering
results.

Our testing revealed that most of the clustering is determined by the
position information of the particles, i.e., the ICL particles are more
clustered in position space ($g=0$), than in velocity space ($g
\rightarrow \infty$).  However, including both the position and
velocity in the clustering ($0 < g \ll \infty$) greatly improved the
quality of the clustering results.  We determined the ``best'' values
for the free parameters, $\ell$, $k$, and $g$, by comparing the
clustered points to features in the simulations that we visually
identified as tidal streams.  For all the following analyses we used
values of $\ell=3.0$, $k=20$, and $g=1.0$.  While our algorithm is
able to detect clustered streams consisting of as few as 21 particles
(the minimum cluster size is $k+1$), we have limited the results to
streams that contain a minimum of 500 particles, corresponding to a
stellar mass of $7.0\times 10^{8} M_{\odot}$, in order to limit our
results to only the largest, most robust streams.

Tidal streams which have been stripped from galaxies during the
process of cluster evolution should consist primarily of particles
from the same parent galaxy.  It would be quite unlikely, given the
high velocity dispersion of cluster galaxies, that two cold streams
from different galaxies would merge to occupy the same phase space.
Thus, in order to find these streams we have run our stream-finding
algorithm independently on each galaxy in the cluster.  The benefit of
this approach is that the stream-finding algorithm is no longer
influenced by the confusing effects of streams from other galaxies or
the diffuse cluster ICL.  That is, the environmental density in which
the stream is located has no effect on its ability to be detected ---
this would be especially problematic at late times in the cluster's
evolution when the interior of the cluster is densely populated by
diffuse ICL.  Thus, while our algorithm defines coherent streams of
ICL clustered in phase space, it is unable to determine how distinct
the streams are from the global ICL distribution.

When the simulation is initialized, each luminous particle is laid
down as a member of a specific galaxy.  However, in order to identify
streams from a galaxy which is the product of merger of two or more
initialized galaxies, we have developed a simple scheme to track the
evolution of galaxies which have merged during the course of the
simulation.  We define a pair of galaxies to have merged when the
separation between the pair becomes less than 5 kpc, and remains less
than this distance for a minimum of 200 Myr.  To calculate the galaxy
separation, we define the position of each galaxy to be the
density-weighted mean position of its luminous particles.  When the
pair merges, we identify the product as a new galaxy, consisting of
all the non-ICL particles that were initialized with the precursor
galaxies; the precursor galaxies themselves are no longer considered.
For ICL particles, however, we identify a particle's parent galaxy as
the galaxy to which the particle belonged at the moment it became ICL.
Thus, it is often the case that we identify an ICL stream as having
originated from a galaxy which has subsequently merged, and can
therefore no longer generate ICL or streams.  This scheme allows us to
properly track the evolution of streams generated prior to a merger,
while also allowing us to identify streams which originate from the
merger product.

\begin{figure*}
\plotone{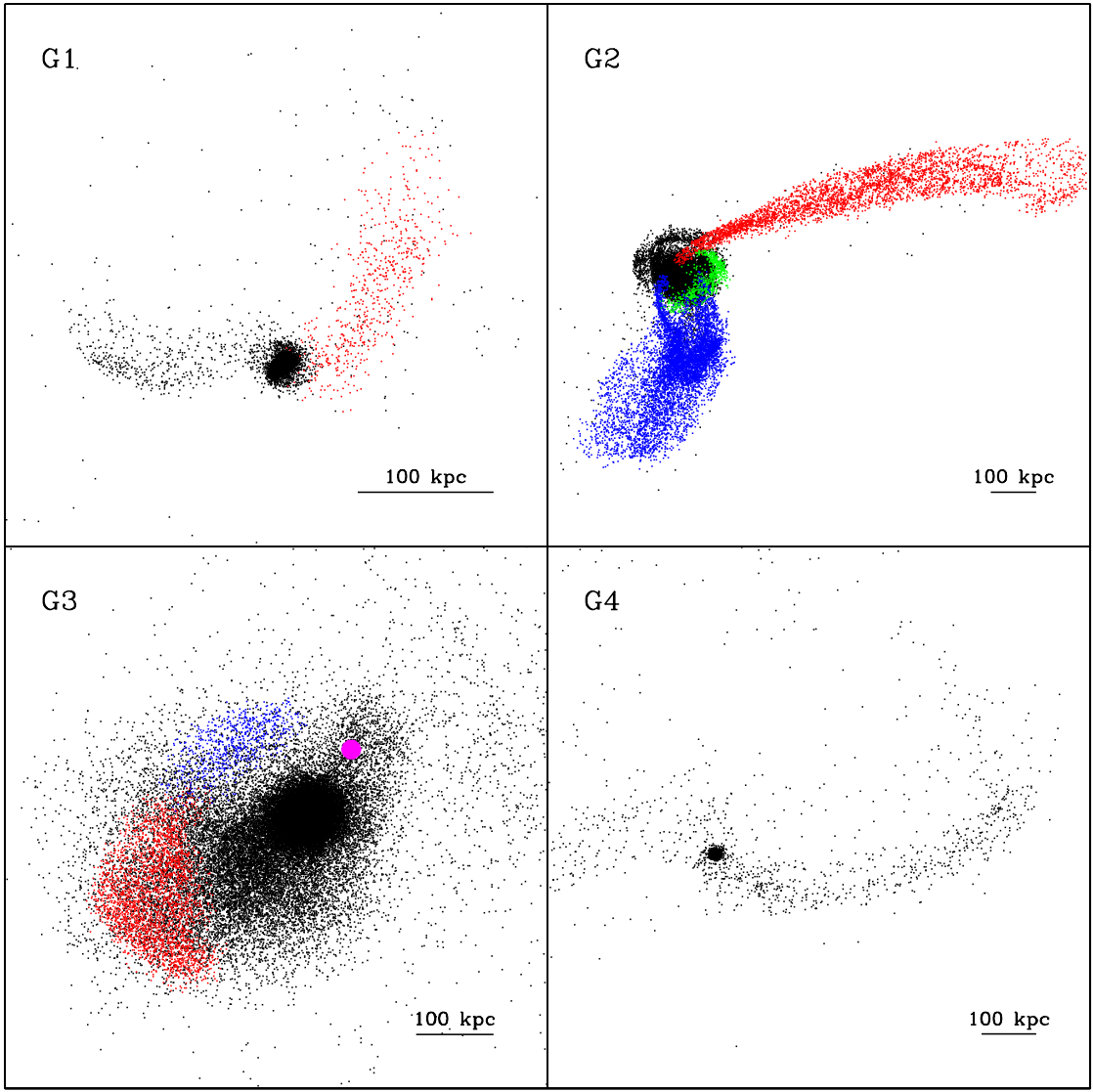}
\caption{
\label{fig:streamplot}
The subfigures each represent a two-dimensional projection of one
galaxy from the 
simulation, labeled G1--G4, respectively.  In black are non-stream
luminous particles from the galaxies; for clarity, only 10\% of these
particles are plotted for galaxies G2 and G3.  In red, blue, and green
are individual streams identified by our algorithm from the galaxies.
The bar in the lower right of each figure shows the physical scale.
The magenta circle in galaxy G3 marks the location of the cD.
}
\end{figure*}

Figure \ref{fig:streamplot} shows some examples of the streams which
we have identified using this algorithm.  The streams display a wide
range of morphologies, from long thin tidal tails (\eg the red stream
from galaxy G2), to more diffuse plume-like structures (\eg the red
and blue streams in galaxy G3).  These streams also span a wide
range in masses from just above our minimum mass limit of 500
particles ($7.0\times 10^{8} M_{\odot}$, \eg, G1 red, G2 green) to
over an order of magnitude larger (\eg G2 blue).  Galaxies G1 and G4
display interesting examples of obvious tidal features which are
present, but our algorithm does not classify them as streams.  The
tidal arm on the left of galaxy G1 is not massive enough to be a
stream; our algorithm finds a group of only $\approx 450$ particles,
just under the minimum stream mass.  The long tidal tail to the right
of galaxy G4, however, is simply too low in density for the particles
to be grouped together by our algorithm.  Section
\ref{sec:streamexamples} contains a more thorough discussion of the
histories of galaxies G1-G3 and the processes which create the streams
seen here.

Because surface brightness is related to the surface mass density (or
column density), we can calculate the surface brightness of material
at a given density at various column depths, assuming a mass to light
ratio.  Following the assumptions of R06, where each particle has a
$V$-band mass-to-light ratio of 5 in solar units, characteristic of an
old stellar population, material at a density of $\rho_{thresh}$ would
have a surface brightness of 30.6 \magsec\ through a 10 kpc column and
28.1 \magsec\ though a 100 kpc column.  While these surface
brightnesses are somewhat lower than the $\mu_V=26.5$ \magsec\ ICL
limit used in R06, in the cluster core we may observe material through
even larger column depths, thus raising the surface brightness.
Observationally, however, this suggests that small streams, with
column depths of $\approx 10$ kpc, may not be detectable through
broadband imaging (\eg Mihos \etal 2005).  Such low surface
brightnesses may only be accessible using discrete stellar tracers,
such as planetary nebulae (\eg Aguerri \etal 2005).  These objects,
however, have the advantage that they can be used to provide kinematic
information (\eg Arnaboldi \etal 2004), which can be useful in identifying
ICL streams.

\section{The Contribution of Streams to the ICL}

One important question about the formation ICL is the physical
mechanism by which it is generated.  If the ICL is created
predominantly through the strong tidal forces caused by close
interactions between pairs of galaxies or between small galaxy groups,
then we would expect to see much ICL initially appear as large tidal
streams.  These types of interactions will generate ICL in short, discrete
bursts, conducive to producing massive, dense tidal streams.
If, however, the dominant mechanism is the slow, gradual stripping of
material as a galaxy orbits in the cluster potential, then we would
not expect to see such features contributing significantly to the
ICL.  The ICL resulting from such interactions would be more evenly
distributed throughout the cluster, with few of the dense
concentrations characteristic of tidal streams.

\begin{figure}
\epsscale{1.2}
\plotone{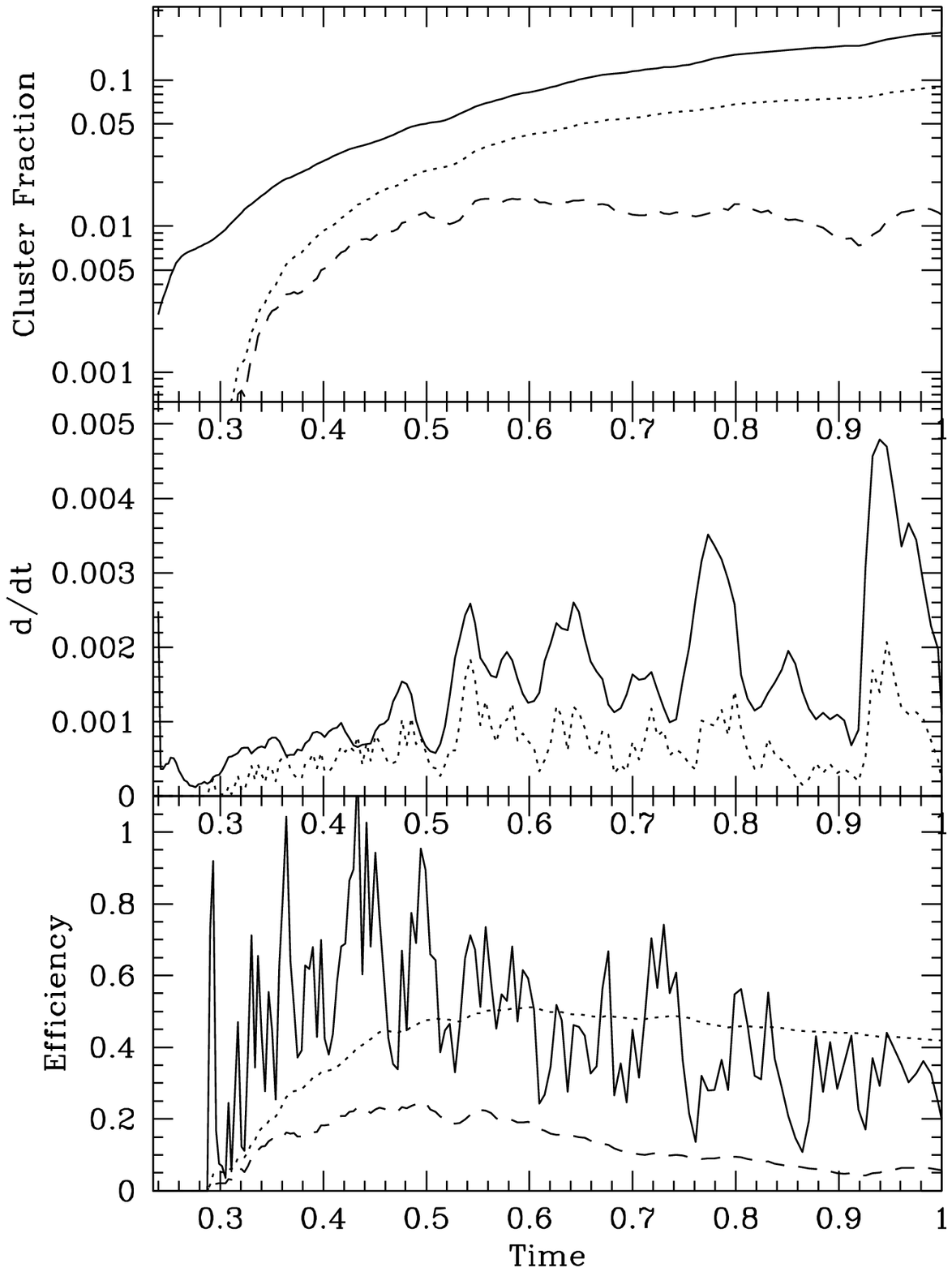}
\caption{
\label{fig:streamfrac} 
Top: the ICL fraction (solid), stream fraction (dashed), and fraction
of particles which have ever been in a stream (dotted) as a function
of time.  Middle: the ICL production rate (solid) and streams
production rate (dotted) as a function of time.  Bottom: the fraction
of the cluster's ICL which is in streams (dashed) or has ever been in
a stream (dotted) as a function of time.  Also shown is the ratio of
the stream production rate to the ICL production rate (solid), as a
function of time.  The time axis is in units of fraction of the age of
the universe (as in R06).  See the text for more details.
}
\end{figure}

The top of Figure \ref{fig:streamfrac} shows the fraction of the
cluster's luminous particles identified as ICL --- the ICL fraction
--- and the fraction of the luminous particles identified as members
of streams --- the stream fraction --- as a function of time.  In
order to focus only on the ICL and streams within the cluster itself,
we have restricted this analysis only to particles which originate
from galaxies which are found within the cluster's virial radius
($R_{200}$) at $z=0$.  The ICL fraction, by definition, is constrained
to be monotonically increasing as a function of time.  At $z=0$ we
find that $21\%$ of the cluster's luminous particles are ICL.  Outside
the cluster virial radius, where galaxies are found predominantly in
small groups with weak potentials, only $5\%$ of the luminous
particles are identified as ICL at $z=0$.

The stream fraction of the cluster rises until $t \approx 0.6$, after
which it begins to decrease somewhat.  This roughly coincides with the
time that the large pre-cluster groups begin to collapse to form a
cluster core, as seen in Figure \ref{fig:cluster_image}.  The decrease
in the stream fraction is a result of the fact that it is dependent on
the rates of two competing processes --- the production of new streams
from tidal interactions and the destruction of old streams as they
dissolve in the cluster potential, to form the diffuse ICL envelope.
In order to disentangle these two competing effects, we have isolated
the production of streams by calculating the fraction of the cluster's
particles which have \emph{ever} been a member of a stream, thereby
ignoring the dissolution of streams.  This analysis is premised on the
simple model of stream formation whereby when a particle is stripped
from a galaxy to become ICL, it does so either as part of a stream or
as diffuse ICL, and after the initial stripping event, or dissolution
of its stream, the particle will become part of the diffuse ICL and no
longer contribute to any stream.  We find these to be good assumptions
for the vast majority of our particles.

To further investigate these trends, the middle of Figure
\ref{fig:streamfrac} shows the time derivatives of the ICL fraction
and the fraction of particles ever in a stream which are equivalent to
the production rates of ICL and streams, respectively.  Before $t
\approx 0.6$, the streams production rate tracks the ICL production
rate quite closely. Thus, early in the cluster's history, when the
cluster core has yet to form and the galaxies are found predominantly
in infalling groups, a large portion of the ICL produced is created in
streams.  However, later in the cluster's history the correlation
between ICL and stream production grows much weaker.  This indicates
that much less of the ICL is being generated in streams as the cluster
core is formed and the cluster potential deepens.

These trends are measured more directly in the bottom of Figure
\ref{fig:streamfrac} where we show the ratio of the production rate of
steams relative to ICL.  Once again, the curve seems to display two
distinct phases of evolution --- before and after $t \approx 0.6$,
when the cluster core begins to form.  Although the ratio of
production rates is quite noisy, it is apparent that early in the
cluster's history, $\approx 60\%$ the ICL produced is generated in
streams, whereas by the end of its' evolution, only $\approx 25\%$ of
the cluster's ICL is generated in streams.  Also shown in the bottom
of Figure \ref{fig:streamfrac} are the ratios of the stream fraction
to ICL fraction, and the fraction of particles which have ever been in
a stream relative to the ICL fraction.  Both curves peak in the $t
\approx 0.5 - 0.6$ range, and by $z=0$ we find that while $\approx
40\%$ of the cluster's ICL was originally generated in streams, only
$\approx 5\%$ of the cluster's ICL is \emph{currently} in a stream.

These data clearly show that early in the cluster's history, when most
galaxies are still living in the pre-cluster group environment, most
ICL is generated in the form of streams, whereas once the massive
cluster potential has formed at later times, streams are created much
less efficiently.  The following sections will look more closely at
the role of the environment in determining the galactic interactions
that generate streams, and the environment's effect on the streams
themselves.

\section{The Galactic Origins of ICL Streams}

In order to understand the processes which create massive ICL streams,
we have closely followed the evolution of individual galaxies within
the cluster, in order to determine when and how they create these
streams.  As galaxies merge, orbit, and evolve in the cluster
potential, they are continuously undergoing an extraordinary range of
interactions that are difficult to quantify in an unambiguous way.
Therefore, here we will concentrate on creating a qualitative
description of the types of interactions which are affecting galaxies
and the effects of these interactions on the production of streams.
Again, ``streams'' here refers to massive, dense, dynamically cold
structures found by the algorithm described in Section \ref{sec:streams}.

\begin{figure}
\epsscale{1.2}
\plotone{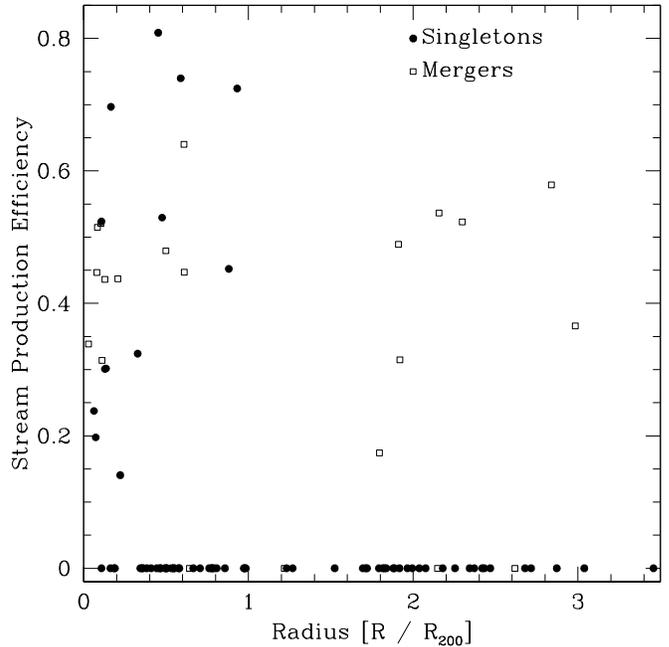}
\caption{
\label{fig:efficiency}
The stream production efficiency, vs. the cluster-centric radius of
each galaxy at $z=0$.  Singleton galaxies (galaxies which have not
undergone a merger) are represented by the solid circles and merged
galaxies are represented with open squares.
}
\end{figure}

Figure \ref{fig:efficiency} shows, for each galaxy, the fraction of
the galaxy's ICL which was produced in streams (the stream production
efficiency), versus the galaxy's cluster-centric radius at $z=0$.  In
general, the galaxies can be divided into two categories, galaxies
ending up within the cluster's virial radius (defined as $R_{200}$),
and those outside.  Outside the virial radius, galaxies that did not
undergo a merging event (singleton galaxies) generated no streams
whatsoever.  Galaxies that did merge show a wide spread in stream
production efficiency, from $\approx 0\%$ to $60\%$.  Within the virial
radius of the cluster, the stream production efficiencies of merged
galaxies are very similar to those of the more distant merged
galaxies.  However, a number of singleton galaxies within the cluster
virial radius have produced streams, with efficiencies ranging from
$\approx 0\%$ to $80\%$.

The differences between these sets of galaxies are due to the
different conditions which the galaxies experience during their
evolution.  The creation of massive streams requires strong tidal
fields which are able to strip large amounts of mass from galaxies
over short periods of time.  For galaxies that undergo mergers, the
most significant tidal interactions are likely to be those between the
merging pairs of galaxies themselves.  These forces are relatively
independent of the local environment beyond the merging pair, and
should thus be similar for galaxies both near and far from the cluster
center.  However, by the large spread in stream production
efficiency seen within the set of merged galaxies, we see that the
details of the merger process can vary considerably and have a
profound effect on the merger's ability to create streams.  Unlike
merged galaxies, singleton galaxies must have another massive
potential --- such as the cD galaxy at the core of the galaxy cluster
--- to interact with in order to create streams.  Galaxies that end up
within the cluster's virial radius are much more likely to have had
such an interaction with a massive galaxy; thus singleton galaxies
outside the cluster virial radius have not produced any streams.  Not
all galaxies within the cluster virial radius have produced streams,
however.  This is due to a number of effects.  First, the strength of
the tidal field is dependent on the gradient of the gravitational
potential, which is much greater in the core of the cluster.  Thus, in
order to have experienced the strong tidal fields required for stream
creation, singleton galaxies must be on highly radial orbits which
bring them very near to the center of the cluster.  Additionally, our
stream detection method is biased against finding streams in galaxies
which are very tightly bound to the cluster, which reside in the
high-density inner regions.  Due to our density-based definition of
ICL, at $z=0$ all of the luminous particles in the inner
$\approx300$ kpc of the cluster are too dense to be classified as
ICL, therefore any galaxy which orbits continuously within this region
cannot create ICL.  Galaxies that reside within this central region
that have produced streams must have created those streams in an
interaction that occurred before they entered the high-density cluster
center.  There are thus two basic scenarios for creating streams, both
of which involve very close interactions between pairs of galaxies:
galaxies in the process of merging; and close interactions between
galaxies that do not merge, often involving the cluster cD.

The scenario outlined above is a useful, but somewhat simplistic,
description of the factors which go into the production of tidal
streams.  The very large scatter seen in the stream production
efficiencies of galaxies with similar merger, location, and orbital
properties is due to the stochastic nature of the high tidal field
events that create streams, and the dependence of the stripping
properties on the specific dynamics of each individual case.  One of
the important systematic dependences affecting stream creation is the
mass of the host galaxy.  Our algorithm detects streams which have a
minimum mass; this mass is equal to up to $9\%$ of our least massive
galaxies.  Thus, low-mass galaxies require a very high fraction of
their stellar mass to be stripped in a single event in order to create
a stream.

\subsection{Examples of Stream Production} \label{sec:streamexamples}
In order to demonstrate some of the aspects of stream production that
have been discussed above, we will show some examples of galaxies on
various orbits in order to demonstrate how the specific circumstances
of each galaxy affect the production of ICL streams.

\subsubsection{Galaxy Interacting with the cD}
\begin{figure*}
\epsscale{0.9}
\plotone{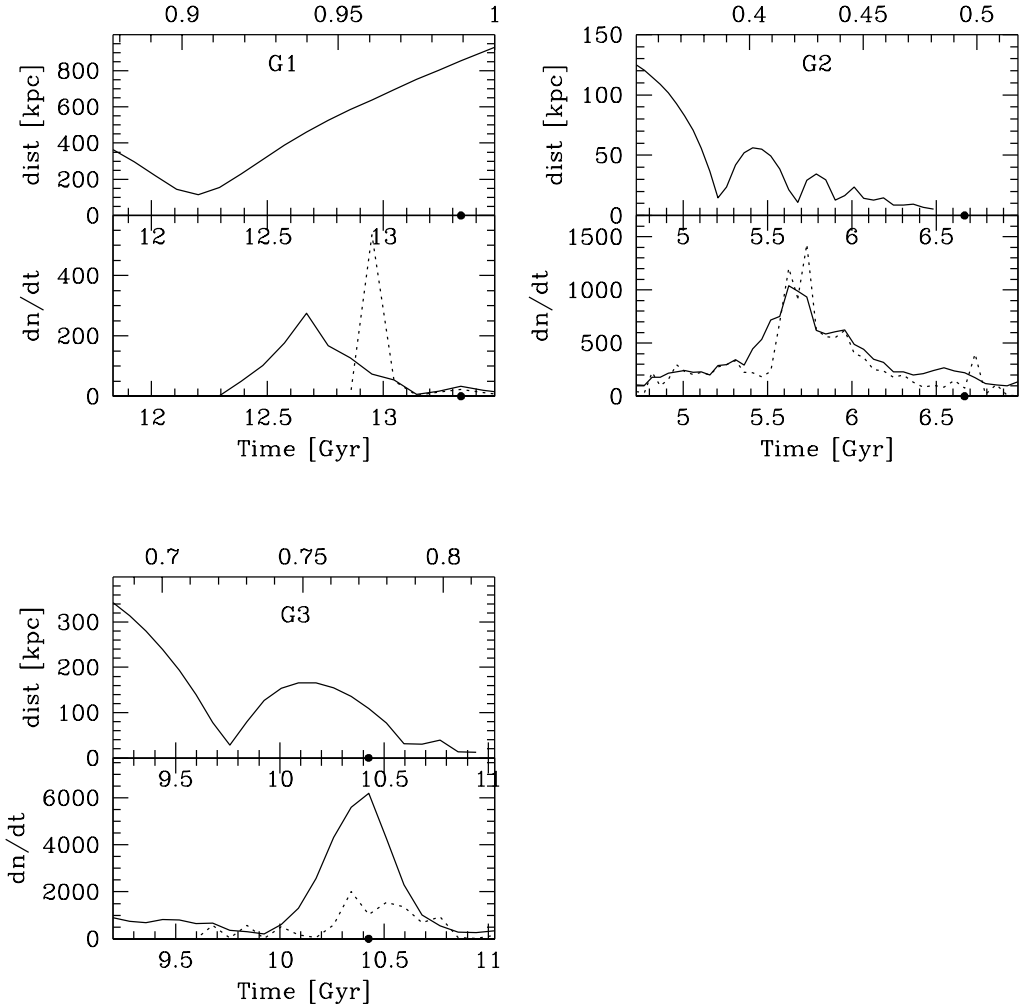}
\caption{
\label{fig:interactions}
The three subfigures represent the three galaxies G1, G2, and G3,
respectively.  Top: the distance between the galaxy in question and
the galaxy with which it is merging (G2 and G3) or the cD (G1).
Bottom: the ICL (solid line) and stream (dotted line) production
rates, for each individual galaxy, in units of particles per timestep.
The bottom time axes show the age of the universe in physical time
units.  However, the top time axis of each plot is fraction of the age
of the universe, as in Figure \ref{fig:streamfrac}.  The dots along
the time axes mark the time when the galaxy is shown in Figure
\ref{fig:streamplot}.
}
\end{figure*}

Our first example galaxy (hereafter named galaxy G1) is a low mass
disk galaxy which interacts with the cluster cD, at late times. This
is a strong, fast interaction ($\approx 650$ \kms) with
closest approach between the two galaxies being $\approx 100$ kpc.
Figure \ref{fig:interactions} shows the distance between the two
galaxies, and both the ICL and stream production rates of G1.  The
production of ICL from this event begins just after the time of
closest approach, and continues to increase for some time afterward
before subsiding.  This is simply a generic consequence of tidal
interactions between galaxies where material is compressed as the
galaxies pass one another and stripped after the passage (\eg
Toomre \& Toomre 1972; Hibbard \& Mihos 1995).  Additionally, there is
no stream production until well after the ICL production has peaked,
and for a short time the stream production rate is higher than the ICL
production rate.  This is simply due to the minimum size requirement
of our stream-finding algorithm; newly created ICL particles cannot
become part of a stream until there are at least 500 ICL particles
which can be clustered together.  Thus, a galaxy may produce ICL for a
short time until a new stream is created, at which time the number of
stream particles must increase by a minimum of 500, making the
stream production rate artificially higher than the ICL production
rate for a short time.

Figure \ref{fig:streamplot} plots the luminous particles of galaxy G1
along with its stream.  This galaxy displays a classic morphology seen
in interactions of disk galaxies orbiting in a potential (\eg Harding
\etal 2001), with both a leading and trailing arm of tidal debris
(again, only one of which we identify as a stream; see Section
\ref{sec:streams}).  We calculate the stream production efficiency of
this galaxy over the time period shown in Figure
\ref{fig:interactions} to be 58\%, as compared with the 45\% stream
production efficiency of the galaxy over the entire simulation time.
The fact that the stream production efficiency due to this interaction
is higher than the galaxy's overall stream production efficiency
indicates that this type of interaction --- a fast, close encounter
with a very massive galaxy --- is particularly efficient at producing
streams, despite the fact that a large fraction of the ICL generated
is part of a cold structure, too low in mass to be a stream.

\subsubsection{Galaxy Merger Outside the Cluster}
Our second example galaxy (hereafter G2) is a massive disk galaxy
which generates streams in a merger with an even more massive galaxy.
This interaction occurs early in the cluster's history, at $t\approx
0.4$, before the cluster core has begun to form.  The galaxy pair is
relatively isolated, $\approx 2$ Mpc from the other large
pre-cluster groups at the time of this interaction, but will fall deep
into the cluster potential by $z=0$.  The mass ratio of the merger is
approximately 4:1, and the more massive galaxy is itself the product
of several previous merging events.

Figure \ref{fig:interactions} shows the distance between the two
galaxies, and the ICL and stream production rates of G2.  As the
galaxies spiral toward one another, galaxy G2 produces a large amount
of ICL, almost all of which is incorporated into streams.  The stream
production efficiency over the time period shown is 86\%.  Due to the
fact that this galaxy merges with another, and therefore no longer
exists as an independent entity after the merger, we cannot calculate
the stream production efficiency of the galaxy over the entire
simulation time.  However, the fact that this event's stream
production efficiency is higher than that seen for any galaxy over the
course of the entire simulation (see Figure \ref{fig:efficiency})
suggests that this type of merging event is particularly efficient at
generating streams.

Figure \ref{fig:streamplot} shows the streams produced by this galaxy,
two of which are very massive (shown in red and blue, respectively)
while the other is of lower mass (shown in green).  A more detailed
analysis of the history of these streams reveals that the two massive
streams were generated by two subsequent close encounters of the
galaxy pair as they spiraled together.  All of these streams show a
generally long, thin morphology similar to that seen in the streams
from galaxy G1.

\subsubsection{Galaxy Merging with the cD}
Our final example galaxy (hereafter G3) is a very massive elliptical,
the product of numerous mergers, that is merging with the even more
massive cluster cD.  Figure \ref{fig:interactions} shows the distance
between G3 and the cD, as well as the ICL and stream production rates
of G3.  The stream production efficiency over the time period shown is
only 30\%, meaning that this type of interaction is much less
efficient at producing streams than the two examples previously
discussed.  This galaxy exists in an extremely complex, dynamically
hot environment, which is not conducive to the creation of dynamically
cold structures.  Additionally, as this galaxy is a massive elliptical
galaxy produced by the merger of numerous galaxies, its stellar
material, even before it is stripped from the galaxy to form ICL, is
likely to be much more dynamically hot than the cold disks which we
examined in the previous examples.

Figure \ref{fig:streamplot} shows galaxy G3 and its streams.  Both of
the streams exist on the left side of the galaxy, as it has just had a
close encounter with the cD, moving from right to left.  This is an
example of the fact that using our ICL definition, ICL, and therefore
streams, cannot be created in the high-density center of the cluster.
Thus the streams are on the low-density side of the galaxy, away from
the cD.  The morphology of these streams, which look like plumes or
shells, is qualitatively very different from the long thin structures
seen in the previous two examples which were much farther from the
cluster center.  The following section will examine more
quantitatively how streams found in diverse environments evolve
differently.

\section{The Life of Streams}
In the previous section we saw that tidal streams that form in
varying circumstances tend to have different morphologies.  We would
expect that these diverse environments should also affect the
evolution of the streams themselves.  In this section we will examine
more quantitatively how streams found in diverse environments evolve
differently, and measure the time it takes for streams to dissolve
into the diffuse ICL background.

In order to follow the growth and decay of an individual stream, we
first have to be able to identify and track the stream as it evolves
through time.  The situation is similar to, but somewhat more complex
than, the case of following dark matter halo assembly histories in a
hierarchical universe (\eg Wechsler \etal 2002), with streams
continually gaining and losing particles, and sometimes merging
together or splitting into multiple pieces.  Unfortunately, the
complexity of the processes makes it prohibitively difficult for us to
simply search through the primary stream catalog and determine each
stream's progenitor(s) and offspring.  We therefore employ an
alternative method, where we follow the evolution of each stream by
eye, in order to determine when the stream reaches its maximum size.
In the majority of cases, the evolution of the stream is easy to
determine, as it simply gains or loses a small fraction of its
particles each timestep.  In situations where multiple streams merge
together, or a single streams splits into multiple pieces, we have to
make a somewhat subjective choice as to the true identity of the
stream.  However, our subsequent analyses are not greatly affected by
these choices, as at this stage we are simply identifying the
timepoint, \tmax\ at which the stream reaches its maximum size.  Once
we have identified \tmax, we select the particles which comprise the
stream at this time and then re-run the clustering algorithm at all
timesteps using only this subset of particles.  Thus, we can follow
the evolution of the stream by tracking a well-defined particle subset.

\begin{figure*}
\epsscale{0.8}
\plotone{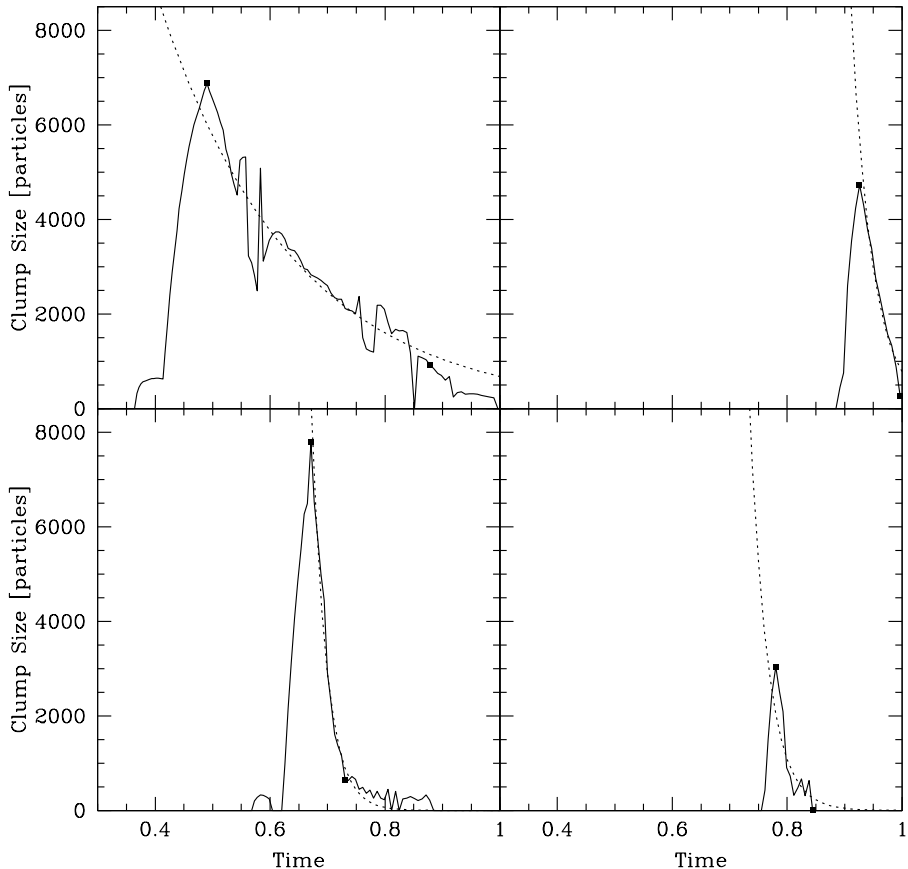}
\caption{
\label{fig:streamhistory}
SOLID LINES: The size of four example streams as a function of time.
DOTTED LINES: The fitted exponential decay of each stream.  SQUARES:
The times between which the decay curves were fit, \tmax\ and the
time when the stream size had decreased by at least 2 $e$-foldings
from the maximum (see text for details).
}
\end{figure*}

Figure \ref{fig:streamhistory} shows the evolution of a few typical
streams from the simulation.  We have measured the size of the streams
in two different ways: the largest group of clustered particles, and
the aggregate size of all clustered groups of particles.  The
measurement method used does not have a strong effect on our basic
conclusions, and all analyses shown measure cluster size using the
largest remaining cluster.

In almost all cases, the streams grow very quickly to their maximum
size and then slowly decay as the stream dissolves into the global ICL
background, or simply disperses until it is to low in density to be
identified by our algorithm.  Unsurprisingly, the rapid initial growth
of the streams indicates that the vast majority of the stream
particles become ICL as a result of a single tidal stripping event
that creates the stream (particles can only be clustered together into
a stream once the particles have become ICL).

More interesting is the decay of the streams as they gradually lose
particles.  There is wide variation in the timescale over which
streams decay, with some decaying quickly and others more slowly.  In
order to quantitatively measure the timescales on which these streams
decay, we have modeled the decay as an exponential decay process, of
the form:
\begin{equation} n \propto e^{-t / \tau} \end{equation}
where $n$ is the size of the stream, $t$ is the time, and $\tau$ is
the characteristic decay time of the stream.  The value of $\tau$ is
obtained by a least squares fit to the data, using the data from time
\tmax\ until the time when the stream size had decreased by at least
two $e$-foldings from the maximum, and stayed below this size for at
least two consecutive timepoints.  Streams which form at late times
which do not decay by at least one $e$-folding by $z=0$ are excluded;
streams which decayed by between one and two $e$-foldings are fitted
between time \tmax\ and $z=0$.  This analysis was restricted to only
the largest streams, which reached a size of at least 2000 particles,
corresponding to a stellar mass of $2.8\times 10^{9}$ \Msun, to ensure
that we could robustly measure their decay.  In addition, we relax our
constraint of having at least 500 particles in a stream for the
descendants of these large streams to ensure we fit over at least two
$e$-foldings in mass decay.  The fits for our example streams are
shown in Figure \ref{fig:streamhistory}.  A small number of streams
have decay times that are not possible to accurately fit, and are
excluded from all further analyses.  These streams generally have
complex evolutionary histories due to the fact that they are created
not from a single stripping event, but from multiple events occurring
in rapid succession, and should therefore be thought of as multiple
independent streams, but are nonetheless aggregated into a single
stream by our clustering algorithm.

\begin{figure}
\epsscale{1.2}
\plotone{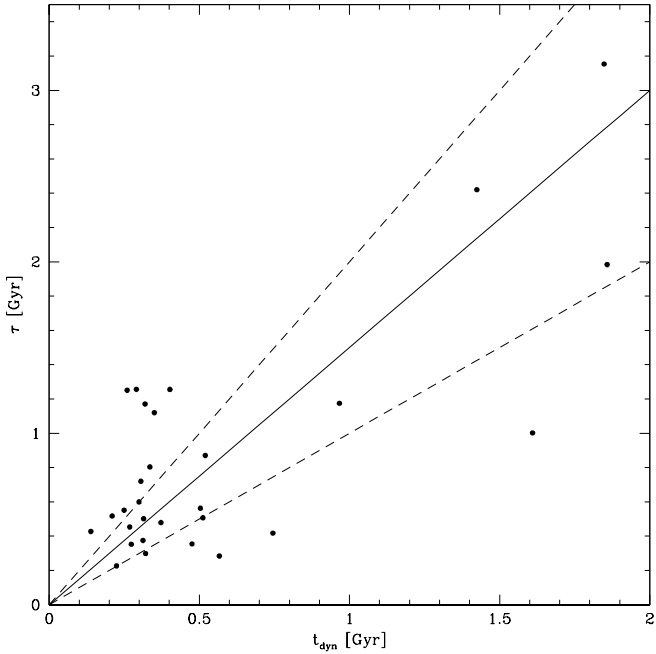}
\caption{
\label{fig:streamdecay}
The decay time, $\tau$, vs. the dynamical time, $t_{dyn}$ for each
stream. The solid line shows relation $\tau = 1.5 t_{dyn}$, while the
dashed lines follow $\tau = t_{dyn}$ and $\tau = 2 t_{dyn}$,
respectively.}
\end{figure}

Figure \ref{fig:streamdecay} shows the decay time, $\tau$, for each
stream versus the dynamical time of the stream in the cluster, given
by
\begin{equation}
  t_{dyn} = \frac{\pi}{2} \sqrt{\frac{r^3}{G M}}
\end{equation}
where $r$ is the stream's cluster-centric radius and $M$ is the mass
enclosed within this radius (Binney \& Tremaine 1987).  The dynamical
time is driven primarily by the stream's cluster-centric radius,
determined to be the distance from the cluster center of mass to the
mean position of the stream particles at time \tmax.  There are
two major sources of uncertainty in the calculation of $t_{dyn}$: the
cluster center of mass is not well defined, especially at early times;
and the streams are often both highly extended and have high
velocities, making their cluster-centric radius somewhat ambiguous.
Despite the large uncertainties that accompany both the measurement of
the decay time and the dynamical time, there clearly is a correlation
between the two, with the decay time being approximately 1.5 times the
dynamical time.  Essentially, we find that streams which are found in
the core of the cluster, with short dynamical times, are quickly
destroyed by the strong, evolving tidal field of the cluster, with
$\tau \la 1$ Gyr.  Streams found farther outside of the cluster are
not subject to these tidal fields, and decay more slowly, with decay
times up to several gigyears in length.  There is a small cluster of points
at $t_{dyn}\approx0.3$ and $\tau\approx1.2$ which lie somewhat above
this relationship.  Each of these streams is formed at
$t\approx0.6-0.7$ in the extremely complex environment of the
collapsing cluster core.  At this time, the cluster potential is
rapidly changing which can significantly bias our determination of
the dynamical timescale.

These data indicate that the evolution of streams is primarily
determined by the external cluster environment where they are found.
In the cluster core, where dynamical times are short and tidal fields
are strong, streams decay rapidly, whereas outside the cluster they
decay more slowly.  However, there are numerous other variables, such
as the velocity and orbit of the stream, the stream morphology, etc.,
which will contribute to stream's evolution.

\section{Summary and Discussion}
In this paper we have identified tidal streams of ICL, calculated
their contribution to the cluster's total ICL, studied the galactic
interactions which give rise to them, and measured their decay
properties.  We have used a density-based definition of ICL, based on
a density threshold below which particles are classified as ICL, in
order to quantify the ICL component using a simple, repeatable
measurement.  Our stream identification algorithm is a
friends-of-friends-type clustering algorithm, which is able to detect
streams using six-dimensional phase space information to find groups
of particles from a common parent galaxy which share similar positions
and velocities.

We find that early in the cluster's history, the majority of ICL is
produced in massive, dynamically cold tidal streams, while at later
times a smaller fraction of ICL is produced in streams.  However, as
in R06, we find that the production of ICL is episodic, suggesting
that it is dominated by discreet events, irrespective of these events
efficiency at forming streams.  Overall, we calculate that $\approx
40\%$ of the cluster's ICL is generated in streams.  We find that
merging pairs of galaxies are relatively efficient at producing
streams because of the strong tidal fields that are created during the
merging process.  However, galaxies which do not undergo a merger are
much less likely to produce streams, and in order to do so must have a
very close encounter with a massive galaxy, often the cluster cD.
Once a stream is created, it begins to lose mass and decay due to the
cluster tidal field.  We calculate the decay time of streams to be
$\approx 1.5$ times the dynamical time of the stream in the cluster.

The stream-finding algorithm we have employed in this paper is
sensitive only to relatively high-density, high-mass
($M\ge7.0\times10^{8}$ \Msun) streams.  ICL substructures which are
too low in mass or density to be identified as streams by our
algorithm, such as the examples seen in Figure \ref{fig:streamplot},
are likely to contain similarly useful information as the streams we
have identified.  However, these low mass or density structures are
also likely to be more difficult to detect observationally.
Similarly, due to our initialization scheme, we are not able to
identify ICL substructures created by galaxies below our minimum
galaxy mass limit ($6 \times 10^{10}$ \Msun).  Although these low mass
galaxies contribute only a small amount of the cluster's luminosity,
their large numbers have the potential to make their ICL substructure
particularly useful in studying the cluster's history and dynamics.

Together, these data suggest a scenario where the dominant mechanism
driving the formation of ICL is highly dependent on the dynamical
state of the cluster.  Early in the cluster's history, when galaxies
are mostly found in small groups which have yet to merge into a single
cluster potential, stream production dominates.  The interactions that
are occurring tend to be close encounters within the group
environment, often involving merging events which are efficient at
generating tidal streams, as with galaxy G2.  Additionally the
relatively weak group potentials found in these environments are
ineffective at dissolving streams, as evidenced by the correlation
between a stream's decay time and its dynamical time in the cluster.
However, as time passes and the cluster potential grows, we see a
different behavior where ICL is less likely to be generated in
large tidal streams.  As we saw with galaxy G3, the mergers of
massive galaxies deep in the cluster potential are much less effective
at generating tidal streams, due to the complex tidal field of the
cluster potential as well as the fact that the stellar material of
these galaxies has already been significantly heated during their
formation.  The streams that result are rapidly dissolved by the
cluster potential.

This scenario is consistent with the findings of R06, which found that
at early times, the morphology of the ICL is dominated by long, linear
features, whereas at late times the ICL became a more diffuse,
amorphous envelope within the cluster.  The long, linear features are
the tidal streams which are created in the group environment early in
the cluster's evolution.  The diffuse envelope seen at late times is
created from tidal stream which have been dissolved by the cluster, as
well as the high fraction of ICL which at late times is not produced
in streams.

We can use this scenario to potentially interpret observational data
of cluster ICL.  In a highly evolved, dynamically old cluster, we
would expect the ICL to show a morphology dominated by a diffuse
envelope, with relatively few streams.  Those streams that do exist
will be relatively short lived, indicating that they were created by
events in the very recent history of the cluster.  In contrast,
a dynamically young cluster dominated by galaxy groups still in the
process of merging is likely to have a higher fraction of its ICL
found in streams.  

The structure of ICL streams found in a galaxy cluster can help us to
understand not only the properties and evolution of the cluster
itself, but also give us information on the galaxies in the cluster.  
We have seen that dynamically hot elliptical galaxies tend
to produce streams with morphologies such as shells or plumes which
are qualitatively different than the long thin tail-like streams
produced by cold disk galaxies.  Therefore, the observed morphology of
a stream is related to the type of galaxy which created it.
Additionally, because streams are created by specific events, such as
close galactic encounters, a galaxy's streams can tell us about its
orbital history.  For example, while galaxy G1 is $\approx 800$ kpc
from the cD at the time it is seen in Figure \ref{fig:streamplot}, its
streams give an indication that it has recently had a major
interaction with one of the cluster's massive galaxies.

Thus, we find that not just the quantity, but the structure, of the
ICL is a useful tool in understanding the evolution and dynamics of
galaxy clusters.  Many of the relevant substructures we have discussed
have very low surface brightnesses and may only be detectable by
incorporating kinematic measurements of the ICL, such as ongoing
planetary nebulae studies.  Such analyses have the potential to
greatly increase our knowledge of the processes that drive galaxy
cluster evolution.

\acknowledgments
C.S.R. appreciates support from the Jason J. Nassau Graduate
Fellowship Fund.  J.C.M. acknowledges research support from the NSF
through grants ASTR 06-07526 and ASTR 07-07793.  This work utilized
the computing resources of the CWRU ITS High Performance Cluster.

\end{document}